\documentclass[fleqn,usenatbib]{mnras}

\usepackage{newtxtext,newtxmath}

\usepackage[T1]{fontenc}

\DeclareRobustCommand{\VAN}[3]{#2}
\let\VANthebibliography\thebibliography
\def\thebibliography{\DeclareRobustCommand{\VAN}[3]{##3}\VANthebibliography}

\usepackage{graphicx}	
\usepackage{amsmath}	

\title[]{The minimum rotation period of millisecond pulsars}

\author[Ertan \& Alpar]{
\"{U}nal Ertan,\thanks{E-mail:unal.ertan@sabanciuniv.edu (\"{U}E)}
M.~Ali Alpar \thanks{E-mail:ali.alpar@sabanciuniv.edu (MAA)}
\\
Faculty of Engineering and Natural Sciences, Sabanc{\i} University, 
Orhanl{\i}, Tuzla, 34956 Istanbul, Turkey}
\date{Accepted XXX. Received YYY; in original form ZZZ}

\pubyear{2021}

\begin{document}
\label{firstpage}
\pagerange{\pageref{firstpage}--\pageref{lastpage}}
\maketitle


\def\be{\begin{equation}}
\def\ee{\end{equation}}
\def\ba{\begin{eqnarray}}
\def\ea{\end{eqnarray}}
\def\m{\mathrm}
\def\d{\partial}
\def\R{\right}
\def\L{\left}
\def\a{\alpha}
\def\acold{\alpha_\mathrm{cold}}
\def\ahot{\alpha_\mathrm{hot}}
\def\Mdotstar{\dot{M}_\ast}
\def\Omegastar{\Omega_\ast}
\def\Omegadot{\dot{\Omega}}

\def\Pmin{P_{\mathrm{min}}}
\def\Peq{P_{\mathrm{eq}}}
\def\OmegaK{\Omega_{\mathrm{K}}}
\def\Mdotin{\dot{M}_{\mathrm{in}}}
\def\Mdotcrit{\dot{M}_{\mathrm{crit}}}
\def\Mdotout{\dot{M}_{\mathrm{out}}}
\def\Mdot{\dot{M}}
\def\Edot{\dot{E}}
\def\Pdot{\dot{P}}
\def\nudot{\dot{\nu}}
\def\Msun{M_{\odot}}

\def\Lin{L_{\mathrm{in}}}
\def\Lcool{L_{\mathrm{cool}}}
\def\Rin{R_{\mathrm{in}}}
\def\rin{r_{\mathrm{in}}}
\def\rxi{r_{\xi}}
\def\rlc{r_{\mathrm{LC}}}
\def\rout{r_{\mathrm{out}}}
\def\rco{r_{\mathrm{co}}}
\def\re{r_{\mathrm{e}}}
\def\Ldisk{L_{\mathrm{disk}}}
\def\Lx{L_{\mathrm{x}}}
\def\Ld{L_{\mathrm{d}}}
\def\Bi{B_{\mathrm{i}}}
\def\Bf{B_{\mathrm{f}}}

\def\Md{M_{\mathrm{d}}} 
\def\NH{N_{\mathrm{H}}}
\def\dEb{\delta E_{\mathrm{burst}}}
\def\dEx{\delta E_{\mathrm{x}}}
\def\Bstar{B_\ast}\def\uff{\upsilon_{\mathrm{ff}}}
\def\Bb{\beta_{\mathrm{b}}}
\def\tint{t_{\mathrm{int}}}
\def\tdiff{t_{\mathrm{diff}}}
\def\r_m{r_{\mathrm{m}}}
\def\rA{r_{\mathrm{A}}}
\def\BA{B_{\mathrm{A}}}
\def\rS{r_{\mathrm{S}}}
\def\rp{r_{\mathrm{p}}}
\def\Tp{T_{\mathrm{p}}}
\def\dMin{\delta M_{\mathrm{in}}}
\def\Rc{\R_{\mathrm{c}}}
\def\Teff{T_{\mathrm{eff}}}
\def\uff{\upsilon_{\mathrm{ff}}}
\def\Tirr{T_{\mathrm{irr}}}
\def\Firr{F_{\mathrm{irr}}}
\def\Tcrit{T_{\mathrm{crit}}}
\def\P0min{P_{0,{\mathrm{min}}}}
\def\Av{A_{\mathrm{V}}}
\def\ah{\alpha_{\mathrm{hot}}}
\def\ac{\alpha_{\mathrm{cold}}}
\def\tc{\tau_{\mathrm{c}}}
\def\p{\propto}
\def\m{\mathrm}
\def\fast{\omega_{\ast}}
\def\Uff{\upsilon_{\mathrm{ff}}}
\def\Ufi{\upsilon_{\fi}}
\def\Ur{\upsilon_{\mathrm{r}}}
\def\UK{\upsilon_{\mathrm{K}}}
\def\Uesc{\upsilon_{\mathrm{esc}}}
\def\Uout{\upsilon_{\mathrm{out}}}
\def\Uphi{\upsilon_{\phi}}
\def\Udiff{\upsilon_{\mathrm{diff}}}
\def\Ure{\upsilon_{\mathrm{r,e}}}
\def\U{\upsilon}
\def\UB{\upsilon_{\mathrm{B}}}
\def\tauB{\tau_{\mathrm{B}}}
\def\hA{h_{\mathrm{A}}}
\def\he{h_{\mathrm{e}}}
\def\cs{c_{\mathrm{s}}}
\def\cse{c_{\mathrm{s,e}}}
\def\hin{h_{\mathrm{in}}}
\def\rhop{\rho^{\prime}}
\def\rhod{\rho_\mathrm{d}}
\def\rhos{\rho_\mathrm{s}}
\def\rhodp{\rho_\mathrm{d}^{\prime}}
\def\rhoe{\rho_\mathrm{e}}
\def\rhoout{\rho_\mathrm{out}}
\def\Alfven{Alfv$\acute{\mathrm{e}}$n~}
\def\Caliskan{\c{C}al{\i}\c{s}kan~}
\def\ql{\textquotedblleft}
\def\qr{\textquotedblright~}

\begin{abstract}
A simple and natural explanation for the minimum period of millisecond pulsars follows from a correlation between the accretion rate and the frozen surface dipole magnetic field resulting from 
Ohmic diffusion through the neutron star crust in initial stages of accretion in low mass X-ray binaries. 
\end{abstract}




\section{Introduction}
\label{sec:intro}
The first millisecond pulsar, PSR J1937+214, was discovered by \citet{Backeretal82}. Immediately after the discovery, 
two groups independently proposed the idea that spin-up by accretion in low mass X-ray binaries leads to 
millisecond equilibrium periods if the surface dipole magnetic fields of the neutron star is in 
the $10^8 \; - \: 10^9 \; G$ range \citep{ACRS82, RadSrini82}. Millisecond pulsars would emerge 
from the epoch of spin-up by accretion on an initial locus in the $P - \dot{P}$ diagram, the \ql spin-up line\qr and proceed to spin-down as a radio pulsar at a rate $\dot{P} \sim 10^{-19}$ s s$^{-1}$. 
Subsequently discovered millisecond pulsars 
all have period derivatives in this range, indicating magnetic fields 
indeed in the $10^8 \; - \: 10^9 \; G$ range. 
The direct confirmation of the 
spin-up by accretion hypothesis came with the discovery of the first X-ray millisecond 
pulsar SAX 1808.4-3658 \citep{WijnandsvdKlis98}. More recently sources exhibiting transitions between 
X-ray and radio pulsar phases have been discovered  \citep{Archibaldetal2009, Papittoetal2013, Bassaetal2014,  Jaodandetal16, PapittodeMartino20}. The shortest millisecond pulsar 
period observed so far from among $\sim$ 400 radio  \citep{catalogue}\footnote{\url{https://www.atnf.csiro.au/research/pulsar/psrcat/}} and 20 X-ray millisecond 
pulsars \citep[see][for a review]{DiSalvoSanna20} is P = 1.4 ms from PSR J1748$-$2446ad \citep{Hesselsetal06}. This is longer 
than the critical break-up period of neutron stars by about a 
factor 3   
for reasonable equations of state  \citep{Heskelletal18}. 
Research on this question has focused on limitations 
to the fastest rotation rates reached by the growth and saturation of neutron 
star modes emitting gravitational radiation 
in the {\em final} stages of spin-up.
We show that a correlation 
between the mass accretion rate and the surface 
dipole magnetic field frozen in the {\em initial} stages of spin-up by accretion explains 
the minimum equilibrium period of millisecond pulsars. 

\section{Magnetic field expulsion from the neutron star}
\label{sec:field expel}
\subsection{Magnetic Field Expulsion from the Superfluid-Superconducting Core}
The fluid cores of neutron stars contain neutrons in the superfluid phase  
and protons in the Type II superconducting phase \citep{Migdal59, GinzburgKirzhnits64, BaymPethickPines69}. 
The rotation of the core super-fluid is enabled by 
an array of quantized Onsager-Feynman vortex lines oriented parallel to the rotation axis while the magnetic flux 
is carried by an array of quantized Abrikosov flux lines parallel to the magnetic axis. External torques acting on 
the neutron star crust are communicated to the core by the interaction of the crust and normal (non-superfluid) 
matter, primarily electrons, with the vortex lines. Under an external spin-down torque the vortex lines move 
away from the rotation axis, thereby reducing the vortex density and achieving spin-down of the core superfluid. The coupling 
between electrons and vortex lines is actually very tight, so the core spin-down lags behind the crust's spin-down 
by minutes or seconds \citep{ALS84}.

The magnetic field in the core proton superconductor can relax by motion of the flux lines whose area 
density determines the mean field. 
The mechanism for magnetic flux expulsion from the core superconductor involves the coupling of 
flux lines and vortex lines. The flux lines and vortex lines inevitably have junctions because of 
their different orientations, and they will get pinned 
to each other because of energy gains at the junctions where they overlap, as \citet{Sauls89} and \citet{Srini89} 
first realized. Spin-down of the neutron star will lead to reduction of the magnetic field in the core 
in proportion to the decrease in the rotation rate, as the vortex lines carry pinned  
flux lines outward \citep{Srini89, SBMT90}. Flux expulsion may be limited at a region of toroidal 
flux lines at the outer core near the boundary with the crust \citep{SideryAlpar09, Erbil14}. 
While the spin-down in the rotation powered pulsar epoch does not achieve significant 
reduction of the magnetic field in the core, the subsequent phase of spin-down by 
accretion from the wind of the detached binary companion does reduce the core field by 
a factor of 100 - 1000 in proportion to the decrease in the rotation rate 
\citep{JahanMiriBhattacharya94, Srini10}. The evolution of the surface dipole moment 
of the neutron star follows the expulsion of the core field by concurrent and subsequent 
processes in the neutron star crust, predominantly by Ohmic diffusion.

\subsection{Ohmic Diffusion of the Magnetic Field through the Neutron Star Crust}

Ohmic diffusion of the magnetic field through the neutron star crust solid depends on the  
conductivity which increases with density and decreases with temperature. 
Accretion increases the temperature, and therefore 
decreases the Ohmic decay timescales and the final value of the surface field. 
In a series of papers Geppert \& Urpin found that the dipole fields of neutron stars can decay by a few orders of magnitude under accretion, reaching final frozen fields in the $10^8$ - $10^9$ G range observed in millisecond pulsars, with values decreasing with increasing mass accretion rate
\citep{GeppertUrphin1994, UrpinGeppert1995, Urpinetal1998}.
Konar and Bhattacharya have made a detailed investigation of 
Ohmic diffusion through the crust in the presence of accretion onto the neutron star 
surface \citep{Konar17, KonBhatt97, KonBhatt99a, KonBhatt99b}. 
However there is also an opposing effect: 
As the accretion rate increases current carrying layers are pushed deeper in the crust, 
to denser regions where the conductivity is much higher. 
This effect wins over, resulting in slower magnetic field 
decay and a correlation between the accretion rate and the final value of the 
surface dipole magnetic field. Konar and Bhattacharya found that the final 
frozen value of the surface dipole moment is reached 
within some 10$^7$ years into the epoch of spin-up by accretion in the LMXB epoch. 
Their work provides a simple and natural explanation of why the surface field does not decay down to zero 
but rather saturates in the observed range of $10^8$ - $10^9$ G. An understanding of the 
minimum rotation period observed among millisecond pulsars follows naturally from this correlation 
between the frozen surface field and accretion rate as we now show. 

\section{Spin-up by accretion and the Minimum Equilibrium Period}
\label{sec:}
The data points shown in Fig. 1 are the results for the frozen field taken from Fig. 4 (lower curve) 
in \citet{Konar17}.  $\Bi$ denotes the initial dipole magnetic field 
prevailing thoughout the crust at the beginning of the spin-up by accretion epoch, 
$\Bf$ is the final, frozen field after the field decay in the initial phases of LMXB  
accretion.  
Its value is at least of the order of the typical surface dipole fields of young neutron 
stars, $\sim 10^{12}$ G, and probably higher.  
The $\Bf/\Bi$ fraction is an increasing function of the accretion rate $\Mdot$, as deeper, higher conductivity regions of the crust determine the final field for higher accretion rates \citep{Konar17}.
Their numerical results can be fit piece-wise with three different power 
laws shown in Fig 1 with the lines A, B, and C.

\begin{figure}
\begin{center}
\vspace{-1.0cm}
\centering
\includegraphics[trim=10cm 76 0 -1cm, width=8.0cm,angle=-90]{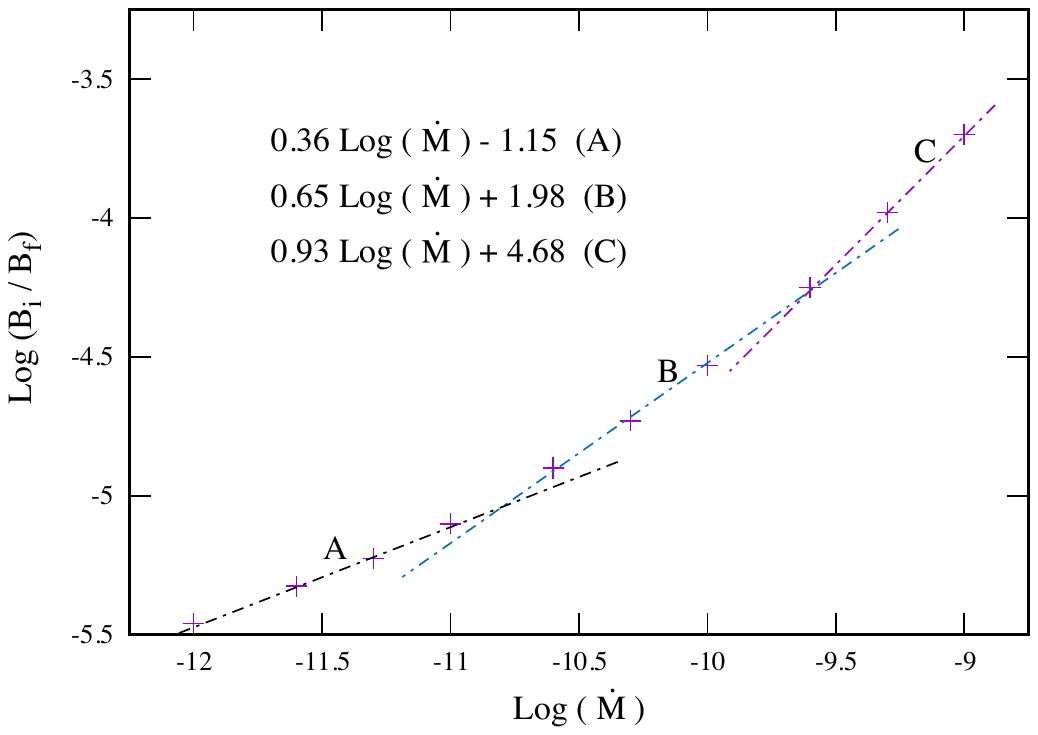}
\vspace{-1.7cm}
\end{center}
\caption{ Fractional decrease in the surface magnetic field as a function of the accretion rate. The lines A, B and C are our piecewise fits to the numerical results, denoted with plus signs, taken from Fig. 4 in \Citet{Konar17} for crustal currents at a density of  $10^{13}$ g cm$^{-3}$ (see the text for details).}
\end{figure}

The equilibrium period $\Peq$ reached after spin-up by accretion depends on the location of the 
inner radius of the accretion disk and on the torques applied on the neutron star by the inner regions 
of the accretion disk.  At the high accretion rates observed in LMXBs the inner disk 
radius is conventionally estimated to be $\rin = \rxi = \xi \rA$,  
where $\rA \simeq (G M)^{-1/7} \mu^{4/7} \Mdot^{-2/7}$ is the \Alfven 
radius, with the value of $\xi$  in the 0.5 - 1 range in many models.
In a recent comprehensive study \citep{Ertan21} showed that $\rin$ can differ 
significantly from $\rxi$ for ranges of $\Mdot$ in the strong-propeller, weak-propeller and spin-up phases. 
The weak-propeller phase of accretion with $\rin \cong \rco$ persists for a 
large range of accretion rates, while the magnitude of the spin-down torque decreases, 
and $\rxi$ approaches $\rco$ with increasing $\Mdot$.  After torque reversal, in the spin-up phase, 
the equilibrium period is reached  
when $\rin \simeq \rco$,   while $\rin$  is close to the \Alfven radius, as in the conventional models. The equilibrium period $\Peq$ is obtained by equating 
the inner disk radius $\rin \simeq \xi \rA$ to the co-rotation radius $\rco = (G M / \Omega^2)^{1/3}$:
\be
\Peq = 2.1  ~   \xi_{0.7}^{3/2}~  M_{1.4}^{2/7}~   \mu_{26}^{6/7}~   \Mdot_{-10}^{-3/7} ~\m{ms} 
\ee
where $\xi_{0.7} = (\xi / 0.7)$,~  $M_{1.4} = (M_\ast / 1.4 \Msun)$,~  
$ \mu_{26} = (\mu / 10^{26}$ G cm$^3$) and $\Mdot_{-10} = (\Mdot / 10^{-10}\Msun $ yr $^{-1})$.  
The equilibrium period, $\Peq$, depends on the dipole moment $\mu$ (corresponding to $\Bf$) 
and $\Mdot$. Using the results in Fig. 1, we can express $\Peq$ as a function of  $\Mdot$ only,  
eliminating $\mu$ using the piece-wise relations A, B and C. 

Using the equations given in Fig. 1 together with Eq. (1), we find that $\Peq$ is 
proportional to $\Mdot^{-0.12}$,  $\Mdot^{0.129}$, and  $\Mdot^{0.37}$ 
for the lines A, B and C respectively. $\Peq$ decreases with increasing $\Mdot$ along A, 
while  it is an increasing function of  $\Mdot$ along the segments B and C, going through a shallow 
minimum. The equilibrium period for each millisecond pulsar is the minimum period it achieves, 
depending on the accretion rate and the correlated surface magnetic field, through 
spin-up by accretion in the LMXB epoch. After the millisecond pulsar emerges as a radio pulsar on the 
'birth line' in the $P\;-\;\Pdot$ diagram it will proceed to spin-down. Having thus noted that
$\Peq$ is the minimum period in the history of any millisecond pulsar, we now proceed to search for the 
minimum equilibrium period $\Pmin$ for the population of all millisecond pulsars.
This $\Pmin$ is obtained with the accretion rate corresponding to the intersection 
point of the lines A and B.  Denoting this rate by $\Mdot_{0}$, we find  $\Mdot_{0} = 10^{-10.8} \Msun$ 
yr$^{-1}  \simeq 10^{15}$ g s$^{-1}$, which corresponds to $\log (\Bf / \Bi) \simeq 5.02$.
For a given solution, $\Bi$ should be consistent with the field strengths of young 
neutron stars, while the $\Bf$ values should be in agreement with the observed range 
of the millisecond pulsars from a few $10^7$ G to above $10^9 $ G. 
Taking $\Bi = 5 \times 10^{12}$ G, and $\xi = 0.7$, we find that 
the $\Mdot$ values in the range $-12 < \log \Mdot  <  -9$ give a dipole 
moment range $0.17 < \mu_{26}< 10$. For this normalization, $\Pmin$ is produced 
with $\Bf \simeq 5 \times 10^{7}$ G. Table 1 shows the $\Pmin$ values obtained with this 
calculation, for each of the numerical data points seen in Fig. 1.    

These results are not sensitive to different $\xi$ and $\Bi$ values and the density layer down to which 
the current loops are pushed. Solutions obtained with different densities can be 
seen in Fig. 4 of \citet{Konar17}. Only one of these model curves is a close 
representation of the accretion rate - frozen field relation. 
Similar ranges of $\Pmin$ values are obtained for a range of $\Bi$ and $\xi$ values with each choice of 
density. The correlation between the frozen field strength and the long-term accretion rate 
establishes a robust barrier to $\Pmin$, with no millisecond pulsars
reaching critical rotation rates.    


\begin{table} 
\caption{ Dipole moments and the minimum periods achieved for the 
accretion rates corresponding to the numerical data seen in Fig.1 
with $\xi = 0.7$ and $\Bi = 5 \times 10^{12}$ G . See the text for details.}  
\begin{center}
\begin{tabular}{c|c|c}

$\mu_{26}$   & $\log \dot{M}$& $\Pmin$ (ms) \\
\hline
 0.17 & -12.0  & 2.00\\
 0.24 & -11.6  & 1.75\\
 0.30 & -11.3  & 1.59 \\ 
 0.40 & -11.0  & 1.51\\
 0.63 & -10.6  & 1.51\\
 0.93 & -10.3  & 1.57 \\ 
1.48 & -10.0  & 1.74\\
 2.81 & -9.6  & 2.03\\
 5.24 & -9.3  & 2.57 \\ 
 9.98 & -9.0  & 3.33 \\ 

\hline
\end{tabular}
\end{center}
\end{table}

\section{Discussion and Conclusions}

We have shown that the minimum equilibrium period of recycled millisecond pulsars can be understood 
naturally in terms of the initial conditions at the beginning of the spin-up by accretion 
in low mass X-ray binaries 
rather than by final conditions when the equilibrium period is defined by a steady state of 
gravitational wave emitting modes of the neutron star. The final \ql frozen\qr value of the surface magnetic 
field and the accretion rate are correlated, because higher accretion rates settle currents to higher density and thereby higher conductivity layers in the crust, as established by 
Konar and Bhattacharya \citep{Konar17, KonBhatt97, KonBhatt99a, KonBhatt99b}. 
As these authors show, the frozen field value is reached at times much shorter 
than the duration of the spin-up by accretion LMXB epoch. 
Thus the surface field value can be taken as an initial condition correlated 
with average mass accretion rate throughout the LMXB epoch. The equilibrium period 
reached when the \Alfven radius and the corotation are equal can be 
expressed as a function of the accretion rate $\Mdot$ alone. Adopting the results 
of Konar \& Bhattacharya we have found that the equilibrium period indeed has a minimum 
at $\Mdot_{0} = 10^{-10.8} \Msun$ yr$^{-1}  \simeq 10^{15}$ g s$^{-1}$, which 
corresponds to $\Bf \simeq 5 \times 10^{7}$ G. This result is not sensitive to parameters 
of the crustal ohmic diffusion. The minimum equilibrium period $\Pmin$ for the millisecond pulsar population is about a factor 3 above the critical neutron star rotation 
period simply because of the frozen field - accretion rate correlation. The shallowness of the 
minimum obtained using the results of Konar \& Bhattacharya also qualitatively accounts 
for the clustering of millisecond pulsar periods near the shorter values. 

\section*{Acknowledgements}
\"{U}E acknowledges research support from
T\"{U}B{\.I}TAK (The Scientific and Technological Research Council of
Turkey) through grant 120F329. 

\section*{DATA AVAILABILITY}

No new data were analysed in support of this paper.

\bibliographystyle{mnras}

\begin{thebibliography}{}

\bibitem[Alpar, Cheng, Ruderman \& Shaham(1982)]{ACRS82}
Alpar M.~A., Cheng A.~F., Ruderman M.~A. \& Shaham J., 1982, \nat, 300, 728
\bibitem[Alpar et al.(1984)]{ALS84}
Alpar M.~A., Langer S.~A., \& Sauls J.~A., 1984b, \apj, 282, 533 
\bibitem[Archibald et al.(2009)]{Archibaldetal2009}Archibald, A. M., Stairs, I. H., Ransom, S. M., et al. 2009, Sci, 324, 1411
\bibitem[Backer et al.(1982)]{Backeretal82}
Backer D.~C., Kulkarni S.~R., Heiles C., Davis M.~M., Goss W.~M.,  1982, \nat, 300, 615
\bibitem[Bassa et al.(2014)]{Bassaetal2014}Bassa, C. G., Patruno, A., Hessels, J. W. T., et al. 2014, MNRAS, 441, 1825
\bibitem[Baym et al.(1969)]{BaymPethickPines69}
Baym G., Pethick C. \& Pines D.,  1969, \nat, 224, 673
\bibitem[Di Salvo \& Sanna(2020)]{DiSalvoSanna20}
Di Salvo T. \& Sanna A. 2020, arXiv e-prints, arXiv:2010.09005
\bibitem[Ertan(2021)]{Ertan21}
Ertan \"{U}, 2021,  MNRAS 500, 2928
\bibitem[Geppert \& Urpin(1994)]{GeppertUrphin1994}
Geppert  U. \& Urpin V., 1994, MNRAS, 271, 490
\bibitem[Ginzburg \&  Kirzhnits(1964)]{GinzburgKirzhnits64}
Ginzburg V.~L. \& Kirzhnits D.~A., 1964,  Zh.Eksperim.Theor.Fiz. 47, 2006
\bibitem[G\"{u}gercino\u{g}lu \&  Alpar(2014)]{Erbil14}
G\"{u}gercino\u{g}lu E.  \& Alpar M.~A., 2014, \apj, 788, L11
\bibitem[Haskell et al.(2018)]{Heskelletal18}
Haskell B., Zdunik J. L.,  Fortin M.,  Bejger M.,  Wijnands R.,  Patruno, A., 2018, A\&A 620, A69
\bibitem[Hessels et al.(2006)]{Hesselsetal06}
Hessels J.W.T., Ransom S.~M., Stairs I.~H., Freire P.C.C., Kaspi V.~M., \& Camilo F., 2006, 
Science 311, 1901
\bibitem[Jahan-Miri \& Bhattacharya(1994)]{JahanMiriBhattacharya94}
Jahan-Miri M. \& Bhattacharya D., 1994, MNRAS 269, 455
\bibitem[Jaodand et al.(2016)]{Jaodandetal16}
Jaodand A., Archibald A. M., Hessels J. W. T., Bogdanov S., D’Angelo C.
R., Caroline R., Patruno A., Bassa C., \& Deller A. T., 2016, ApJ, 830, 122
\bibitem[Konar(2017)]{Konar17}
Konar S., 2017, J. Astrophys. Astr. 38, 47
\bibitem[Konar\& Bhattacharya(1997)]{KonBhatt97}
Konar S. \& Bhattacharya D., 1997, MNRAS 284, 311
\bibitem[Konar\& Bhattacharya(1999a)]{KonBhatt99a}
Konar S. \& Bhattacharya D., 1999a, MNRAS, 303, 588
\bibitem[Konar\& Bhattacharya(1999b)]{KonBhatt99b}
Konar S. \& Bhattacharya D., 1999b, MNRAS, 308, 795
\bibitem[Manchester et al.(1993)]{catalogue}
Manchester R. N., Hobbs G. B., Teoh A., Hobbs M., 2005, ApJ, 129, 1993
\bibitem[Migdal(1959)]{Migdal59}
Migdal A., 1959, Nucl.Phys, 13, 655
\bibitem[Radhakrishnan \& Srinivasan(1982)]{RadSrini82}
Radhakrishnan R. \& Srinivasan G., 1982, Curr.Sci, 51, 1096
\bibitem[Papitto et al.(2013)]{Papittoetal2013}Papitto A., Ferrigno C., Bozzo E., et al. 2013, Nature, 501, 517
\bibitem[Papitto \& de Martino(1999a)]{PapittodeMartino20}Papitto A. \& de Martino D., 2020, arXiv e-prints, arXiv:2010.09060
\bibitem[Sauls(1989)]{Sauls89}
Sauls J.~A., 1989, in "Timing Neutron Stars", 
{\"O}gelman, H., van den Heuvel, E.P.J. (eds.) Kluwer, Dordrecht
\bibitem[Sidery \& Alpar(2009)]{SideryAlpar09} 
Sidery T. \& Alpar M.~A., 2009, MNRAS, 400, 1859
\bibitem[Srinivasan(1989)]{Srini89}
Srinivasan G., 1989, A\&A 1, 209
\bibitem[Srinivasan et al.(1990)]{SBMT90}
Srinivasan G., Bhattacharya D., Muslimov A.~G. \& Tsygan A.~I., 1990, Curr.Sci. 59, 31
\bibitem[Srinivasan(2010)]{Srini10}
Srinivasan G., 2010, New Astronomy Reviews 54, 93
\bibitem[Urpin \& Geppert (1995)]{UrpinGeppert1995}
Urpin V., \& Geppert U., 1995, MNRAS, 275, 117
\bibitem[Urpin et al.(1998)]{Urpinetal1998}
Urpin V., Geppert U., \& Konenkov D.,  1998, A\&A, 331, 244 
\bibitem[Wijnands \& van der Klis(1998)]{WijnandsvdKlis98}
Wijnands R. \& van der Klis M., \ 1998, \nat, 394, 344
\end{thebibliography}

\bsp	
\label{lastpage}
\end{document}